# FAIR-USE4OS: Guidelines for Creating Impactful Open-Source Software

**Raphael Sonabend**[1,2] *
**Hugo Gruson**[3]
**Leo Wolansky**[4]
**Agnes Kiragga**[1,5]
**Daniel S. Katz**[6]

1. OSPO Now, UK
2. Imperial College London, UK
3. data.org, US
4. Scale AI, US
5. APHRC, Kenya
6. University of Illinois Urbana-Champaign, US

*Correspondence: raphaelsonabend@gmail.com

## Abstract

This paper extends the FAIR (Findable, Accessible, Interoperable, Reusable) guidelines to provide criteria for assessing if software conforms to best practices in open source. By adding 'USE' (User-Centered, Sustainable, Equitable), software development can adhere to open source best practice by incorporating user-input early on, ensuring front-end designs are accessible to all possible stakeholders, and planning long-term sustainability alongside software design. The FAIR-USE4OS guidelines will allow funders and researchers to more effectively evaluate and plan open source software projects. There is good evidence of funders increasingly mandating that all funded research software is open source; however, even under the FAIR guidelines, this could simply mean software released on public repositories with a Zenodo DOI. By creatingFAIR-USE software, best practice can be demonstrated from the very beginning of the design process and the software has the greatest chance of success by being impactful.

## Author Summary

This research builds on the FAIR principles to ensure research software adheres to open source development best practice, which includes community engagement and early planning for long-term sustainability. By creating guidelines ('FAIR-USE4OS') that can be followed, funders and researchers are in a stronger position to evaluate and create research software with maximal chance of success. This research is important as open-source software that is not 'FAIR-USE' has a lower probability of long-term impact. These guidelines will help benefit and impact society once they are widely accepted by researchers and funders, which could happen within a relatively short time period given good evidence that funders are actively including and updating open source policies, which directly impact upon how research is conducted.

## Introduction

In the simplest definition, open-source software is code released under a license that allows users to use, adapt and reshare the software in a way that is far less restrictive than traditional closed-source, proprietary software. Open-source licenses (often defined as those approved by the Open Source Initiative (OSI) [1]) vary in how restrictive they are, particularly with respect to adaptation, redistribution, sublicensing, and more. Open-source software (OSS) is often conflated with 'free/libre and open source software' (FOSS/FLOSS), a social movement focused on the freedom to use software however one pleases. In recent times, the term 'open source' has also started to refer to a movement of principles that focuses on openness, transparency, and accessibility, other terms for this movement include "open science", "open community", and "open development". In this paper, we define 'FAIR-USE' software, which encapsulates all these movements and refers to software that not only has an OSI license but also conforms to best practice principles thus being genuinely welcoming to all users, including contributing developers.

## Results/Discussion

### Defining the Problem

Thanks to recent efforts from funders to cost-in specialist roles for developing scientific software [2], we now find developers (usually research software engineers) being paid to create high-quality code alongside domain-experts/principle investigators costed to develop methodology; increasingly community managers are also included on grants to develop and engage communities of practice around the software. Whilst funders will often mandate the resulting code is open-source in order to maximize impact [3][4], researchers are typically not funded to ensure their software is genuinely usable by end-users. This is visualized in the top row of Figure 1, which demonstrates the journey to increasingly FAIR-USEprojects. We believe that software projects often end at the 'open-source code' stage before fully aligning with FAIR-USE principles, with the consequence that the software does not reach its full potential in terms of utility and is less impactful than it could be. As our definition of FAIR-USE focuses on *principles*, we see no reason for 'immature' software not to be considered FAIR-USE as long as it continues on this path with increasing maturity, which is reflected in the final, red 'Iteration' boxes in Figure 1.

In this paper we introduce the FAIR-USE4OS guidelines (pronounced 'Fair use for OS'), which build on the FAIR principles to highlight this gap between open-source software (OSS) and FAIR-USE software, to help funders and developers ensure that adequate resourcing is provided to ensure software can adhere to open source best practice.

Before going into detail on the new guidelines, we want to emphasize that code is open source *only* when made available under an open-source license, it is not sufficient to just openly publish code on a public repository such as GitHub or GitLab without an explicit open-source license. Ideally, the license should be approved by the Open Source Initiative (OSI) [1]. The OSI definition of 'open source' includes 10 clauses to determine if code is 'open source', the most important of which are that the code can be redistributed for free, the source code must be obtainable (possibly at a reasonable cost), and derivatives of the code can be made and distributed possibly under the same license. Licenses fall under several categories: the most commonly used are 'permissive', which mean code can be freely distributed and modified in any way and redistributed under any license as long as the original source is credited; 'weak copyleft', which adds the requirement that redistribution (and possibly modified) code is licensed under the same terms as the original; and 'strong copyleft', which adds that any code linked or combined with the original must also use the same (or compatible) license. For research software, we recommend the use of permissive or weak copyleft licenses where possible to ensure future researchers can build on the software with minimal disruption. We do not explicitly recommend use of strong copyleft licenses, due to the restrictions they place on onward development and the risk (and frequent occurrence) of developers using incompatible licensing (such as more permissive licenses) in downstream dependencies.

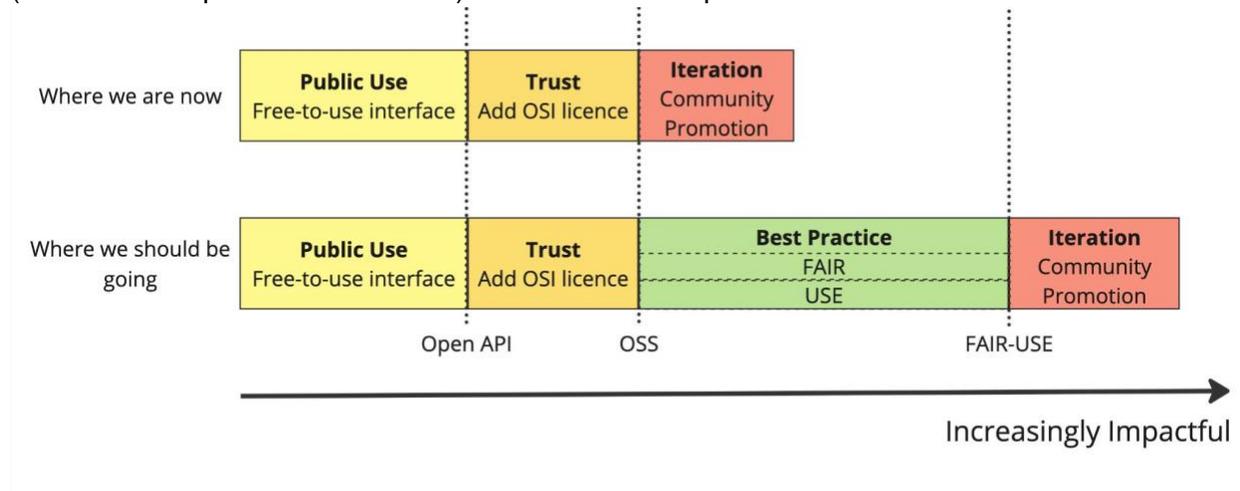

Figure 1: Top row: simplified version of the past and current landscape for funding research software, noting that usage of FAIR guidelines and mandating clear sustainability ideas have increasingly appeared. The first, yellow box represents the first stage in creating FAIR-USEsoftware, providing a free to use public interface. Software in this case may erroneously be viewed as 'open'. However, without an OSI-approved license, the code is not open source. Common examples of this are websites that provide APIs for developer use. The second, orange box represents adding an OSI license and releasing all (or most) of the codebase for free, we refer to this as the 'trust' stage as making code open-source depends on developers and researchers trusting this will only improve usage and not lead

to 'competitors stealing' code (a commonly cited reason not to make code open source). The third, red box represents open-source software that has been released and promoted to the intended community. Software is never truly 'finished' hence this last stage is referred to as 'iteration', signifying the importance of constant user feedback for updates. In the top row, the 'iteration' stage is reached before software is FAIR-USE. Bottom row: where we believe the funding landscape should be heading with more emphasis on FAIR-USE4OS as described in this paper. The addition of FAIR and USE allows projects to move from OSS to FAIR-USE (third, green box). The arrow from left to right indicates that each of these steps makes a project increasingly likely to be impactful. Skipping the green box means software may not be sufficiently FAIR-USE to be high impact; more focus on FAIR-USE4OS guidelines can fill this gap.

## FAIR-USE4OS

We begin by outlining our full criteria (Table 1), which we then break down into distinct sections, focused on USE and FAIR. We include '4OS' in the name of the guidelines to highlight that our expanded FAIR guidelines are for software (explicitly, open-source software) and not data. We only recap the FAIR criteria, which are fully defined for research software in Barker et al. (2022) [5].

Table 1: The FAIR-USE4OS guidelines. The high-level headings of FAIR are copied from Barker et al. (2022) [5].

| **F**indable | Software, and its associated metadata, is easy for both humans and machines to find. |
|---|---|
| **A**ccessible | Software, and its metadata, is retrievable via standardized protocols. |
| **I**nteroperable | Software interoperates with other software by exchanging data and/or metadata, and/or through interaction via application programming interfaces (APIs), described through standards. |
| **R**eusable | Software is both usable (can be executed) and reusable (can be understood, modified, built upon, or incorporated into other software). |
| **U**ser-centered | Stakeholders of the software have been involved in co-design, development, and testing of the software to ensure it genuinely meets their key user needs, including precise public-facing documentation to minimize the barrier to entry. |
| **S**ustainable | Long-term sustainability plans are in place for the codebase, community, and finances. |
| **E**quitable | Software can meet the needs of all intended users including localisation, low-bandwidth, and low-resource options where needed. |

The USE guidelines build on FAIR by explicitly stating some of the principles in FAIR that are included in spirit but not explicitly. For example, the FAIR4RS principles only require code to have a clear license, but do not require software to be open-source [6], whereas FAIR-USE4OS can only be applied to open-source software. We begin by introducing USE, then recap FAIR and highlight the similarities and differences between the two.

## USE: User-centered, Sustainable, Equitable

The core of this paper rests on the belief that open-source software cannot be impactful if it is not usable for all intended users. This may seem obvious, but historically funders have focused on mandating software be made open source and expect the software to go on to have real-world impact, but with no midpoint to validate if the software is in fact usable. This means that either research software engineers or community managers are expected to perform this duty, despite people in these roles often not having the required time or skills [7], or the green box in Figure 1 is simply ignored (which is the current problem that we are facing). Therefore, resulting software has the *potential* to solve real-world problems, but lacks the interface to make this possible. In this section we advocate for the 'USE' (**U**ser-centered, **S**ustainable, **E**quitable) reporting guidelines outlined below to ensure software is genuinely usable and welcoming to all users.

### U: User-centered

User-centered software is essential to ensure that the resulting software solves a genuine, real-world need. The term 'user' does not refer to only detached end-users, but also the package developers, developers of downstream packages, and other stakeholders who may directly or indirectly interact with the software. If resources are available, then user-centered design should be considered from the start of a project to rigorously assess the usability and utility of the software throughout the design process. There are many formalized methods for user-centered design, which are increasingly accessible with collaborative, online software. In practice, it is rare for projects to have sufficient resources for end-to-end user-centered design and development, so instead, we encourage co-design with suitable experts at regular stages throughout the project. These experts could be a small group of potential end-users from the target users, community leaders who can represent the views of the intended users, or a project steering committee. The experts should be provided the chance to give feedback on the intended purpose and usability of the software to ensure it will genuinely solve the real-world problems that they/they're community are facing, and to highlight any potential downsides to designs. As well as providing design input, we recommend early release of a prototype that users can trial in order to identify bugs or design flaws that can be fed back to developers. Once a prototype is live, a community manager may be best placed to facilitate feedback from users to reduce the burden on the developer.

Whilst often seen as an administrative burden, documenting code is a useful method to center users in the code development process. Documentation increases usability by making the codebase more transparent and tells users and other developers how to interact with the code. Open source software depends on its community and high-quality documentation lowers the barrier to entry for all potential end-users. Regularly documenting new or updated code, provides the opportunity for developers to consider the software from the user-perspective. To guide this process, as a general rule, all public-facing functionality should be consistently documented, at the very least with function inputs (parameters) and outputs (values returned) clearly defined. This should also include descriptions of core functionality, such as key methodology that is applied with citations where appropriate. Documentation should be understandable by non-technical audiences with as little jargon as possible. As well as documented public-facing functionality, code commenting should be used for *any* code that is not intuitive; for example, long functions should have in-line code comments to keep track of progress through the function, and all functions should include very basic commenting at the start to outline inputs, outputs, and functionality. The amount of documentation is a subjective trade-off that may be co-defined with developers and users. There may be cases where robust frameworks such as Diataxis [8] are appropriate, however this is not a requirement for all software and documenting should not be done for the sake of it.

## S: Sustainability

To achieve full impact, software must be sustained over time and correctly marketed. In terms of sustainability, there are three primary areas that must be sustained: community and users, funding, and the codebase.

Without an active and engaged community of users, the software fails to solve a real-world problem and will not be impactful. We define a 'community' here as a 'community of practice': a group of people united around a common theme, in this case the software. This community could be a handful of users or thousands of users, depending on the needs of the software and the problem it is solving. Communities and users should not be forced to use software for the sake of it and if, over time, it becomes apparent that users are dropping off as the software is no longer required, then sunsetting (retiring) software is a natural endpoint. On the other hand, if the software continues to be relevant, then community managers or equivalent should be engaged to ensure the community is active, provides feedback and reports bugs - there is positive evidence that funders are increasingly providing funding for development and hiring of community manager positions for this purpose (e.g., [10]).

Maintaining and sustaining code is simplest when a community is most active, however even without an active community, code should be routinely updated and maintained. Bug fixing is required to ensure the software remains trustworthy, and implementing new features is required to ensure the software remains relevant. Automated testing of code is best practice and should be implemented by developers whenever possible. Testing of code is often formalized as verification and validation (V&V), which are procedures for checking that software correctly meets its intended purpose. By considering sustainability alongside the User-centered guideline discussed above, developers should also consider how stakeholder engagement and end-user testing should be incorporated into ongoing testing and maintenance.

Finally, funding for software can be acquired via grant funding, creating software as a service, offering tiered subscriptions (e.g., for quicker bug fixing), dual-licensing, and other models - there is no turnkey solution for financial sustainability and all software will need a bespoke solution. Note that acquiring future grant funding is usually only possible with demonstration of existing success and therefore community and code sustainability must be viewed as equally important as financial sustainability.

## E: Equitable

Software is equitable if it can meet the needs of all its intended users. In contrast to the User-centered guideline above, which focuses on ensuring software solves a real-world problem, this guideline focuses on ensuring that all intended users of the software will find it equally usable and accessible. As a principle, equitable software should be designed to benefit users across as many contexts as possible, but at a bare minimum, it should verifiably serve the needs of those it is designed to benefit. Part of being equitable is therefore spending the time to determine and understand the target users and their varying needs and contexts. Once understood, software must be designed to be usable and accessible by all of those users. For example, a software package might be technically brilliant and tackle essential real-world problems, but if many users cannot use the software because they are unable to learn complicated new syntax, then the software has not been developed equitably. To be equitable, software may need to be localized (e.g., using tools such as CrowdIn) and discounts may be required for paid-for software (for example open-source code with a commercial front-end) - as discussed below, resources will limit how feasible it is to localize and discount all solutions, and developers and community managers should work together to identify minimum requirements to remain equitable (e.g., limiting localization to documentation translation). Note that we do not see commercial front-ends as a violation of FAIR-USE principles as long as the underlying code is open-source - in fact partial commercialisation is an important part of ensuring FAIR-USE software remains sustainable (discussed in the next section).

In addition to these requirements, equitable code should also be findable (i.e., not simply dumped in a public software repository but also added to professional websites and other public resources), include clear instructions for installation and usage (this is a subset of documentation), and can be tailored to intended users with low-bandwidth (e.g., offline solutions are possible). Whilst some software will require high performance computing (HPC) and/or cloud access in order to be run, these resources should only be utilized when strictly required (and not for the sake of it); in these cases, equitable procedures must be put in place to accommodate users without access to these resources, which might mean providing lower-complexity models/software that does not require HPC/cloud or paying for access to these resources for users who would not otherwise be able to afford it. Developers are not expected to plan for every possible use case on their own, and there will be many situations that are hard to anticipate in advance without understanding individual need; therefore, developers should work with community managers to ensure software is iterated as required (within reasonable limits). To assess if software is equitable, developers should identify key stakeholders to understand potential use cases and in each of these use-cases, the software should be verified to be equally usable and accessible, with adjustments iteratively made as required until equity is reached. Using wikis is one method to create guides for different use-cases so users can easily identify their expected journey.

Note that we have not discussed 'ethical' software above as we view this as being part of the greater funding proposal and not a metric to judge software by. However, we would expect an ethical proposal for research software to, at a minimum, demonstrate how the software will be equitable.

## FAIR: Findable, Accessible, Interoperable, Reusable

USE builds on the FAIR [9] and FAIR for research software (FAIR4RS) [5] guidelines. By definition, FAIR research software is easy for humans and machines to find, retrievable via standardized protocols, interoperable with other software, and usable and reusable.

In the simplest case, software may be Findable and Accessible if it is released on a public repository (such as on GitHub, GitLab, Bitbucket or other similar platform), has a DOI from Zenodo or another provider, and includes descriptive metadata. Interoperability ensures that software can interact autonomously with other software over time without manual intervention. This is important for impact as long-term success for open-source software can often mean maintaining the current codebase without implementation of new features in order to be a dependency of newer software - this highlights an important nuance in the definition of 'Impact', some pieces of software will not have direct impact by solving a real-world scientific problem but will have huge indirect impact, for example open-source compilers.

The key difference between FAIR and USE is that the former is focused on the software as a product or entity in that FAIR software can be queried, accessed, and executed with computational commands but does not explicitly consider how a human interacts with the software beyond these computational queries (though proponents of FAIR would likely say this is captured in spirit of the FAIR guidelines), while on the other hand, USE focuses on the user-journey when interacting with the software and its community. Together, USE software that passes the FAIR guidelines (FAIR-USE), will be human- and machine-usable, ensuring that the software can be found, accessed, and incorporated into other software routines (FAIR), as well as being intuitive and 'easy' to use for all intended users regardless of their background and resources (USE).

We believe FAIR-USE (Figure 2) software has a greater chance of impact than software that is simply open source. To ensure impactfulness can be reached as quickly as possible, funders should share responsibility for ensuring that the correct community (i.e., the intended users) for the software is found and there is positive evidence that funders are already taking these steps [11].

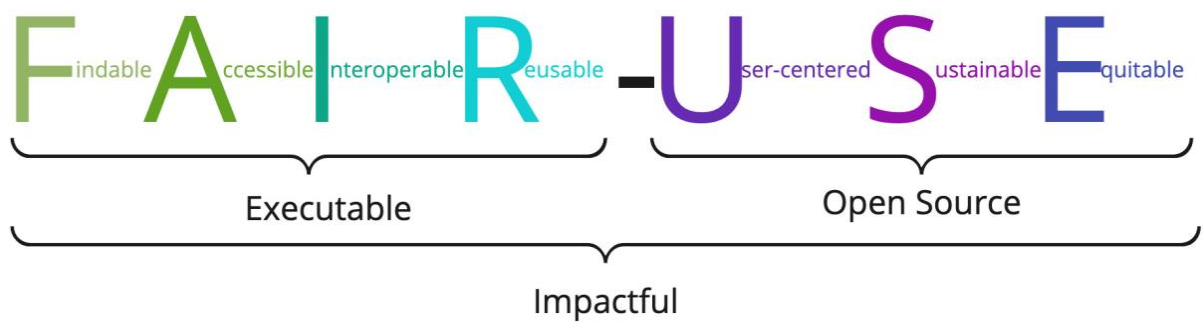

Figure 2: Summarizing the FAIR-USE4OS guidelines. FAIR software can be executed by a machine but may not necessarily be open-source software. USE software is open source and user-friendly. Together, FAIR-USE software has the greatest chance of impact over time.

## Discussion

In this paper we identified a gap in how research software is currently funded, in which funders expect impactfulness to follow from creating high-quality open-source code, but in fact, the resourcing required to ensure open source best practice is greater than usually anticipated. To evaluate this gap, we defined requirements by augmenting the FAIR guidelines with criteria that can be used to assess if best practice is adhered to: **U**ser-centered, **S**ustainable, **E**quitable (USE), creating the FAIR-USE4OS guidelines.

As a final comment, we note that Eaves et al. 2022 [12] distinguish open source software (OSS) from Digital Public Goods (DPGs) by stating the requirements that DPGs must be open-source as well as intentionally defined, equitable in use, sustainable, and scalable. These requirements overlap nicely with the guidelines in this paper, so that the step from FAIR-USE4OS software to DPG becomes one primarily of governance. Whilst the sustainability requirement in FAIR-USE4OS may cover governance (and funders will often insist on this), this is rarely the formalized governance required for DPGs managed or developed at a governmental or NGO level.

We hope that this paper helps creators of research software, as well as funders, to use the FAIR-USE4OS guidelines to write and assess funding applications respectively. To ensure software can be as impactful as possible, we call on funders to provide more resourcing specifically to tackle user-centered front-end design to create FAIR-USE software.

## Author contributions

Conceptualization: RS and AK. Writing - Original Draft: RS, HG, DSK. Writing - Review and Editing: RS, HG, DSK, LW. Visualization: RS, HG, DSK.

## Acknowledgments

DSK was partially supported by the Alfred P. Sloan Foundation.

## Competing Interests

None.